\documentclass[a4paper,fleqn,usenatbib]{mnras}
\usepackage[T1]{fontenc}
\usepackage{ae,aecompl}
\usepackage{graphicx}	
\usepackage{amsmath}
\usepackage{amssymb}

\title[The population  of WD+MS in the SDSS DR12] {The population 
   of white dwarf-main sequence binaries in the SDSS~DR~12}

\author[R. Cojocaru et al.]{R. Cojocaru$^{1,2}$, A. Rebassa-Mansergas$^{1,2}$,
  S.  Torres$^{1,2}$, E. Garc\'ia-Berro$^{1,2}$\thanks{Email;
  enrique.garcia-berro@upc.edu}\\
$^{1}$Departament de F\'\i sica, Universitat Polit\`ecnica
  de Catalunya, c/Esteve Terrades 5, 08860 Castelldefels,
  Spain\\
$^{2}$Institut d'Estudis Espacials de Catalunya, Ed. Nexus-201, 
  c/Gran Capit\`a 2-4, 08034 Barcelona, Spain}

\date{Accepted. Received; in original form}

\pubyear{2016}

\begin{document}
\label{firstpage}
\pagerange{\pageref{firstpage}--\pageref{lastpage}}
\maketitle

\begin{abstract}
We  present  a  Monte  Carlo   population  synthesis  study  of  white
dwarf-main sequence  (WD+MS) binaries  in the  Galactic disk  aimed at
reproducing the ensemble properties  of the entire population observed
by  the  Sloan  Digital  Sky  Survey  (SDSS)  Data  Release  12.   Our
simulations take into  account all known observational  biases and use
the most  up-to-date stellar evolutionary  models.  This allows  us to
perform   a  sound   comparison  between   the  simulations   and  the
observational data.  We find that  the properties of the simulated and
observed parameter  distributions agree best when  assuming low values
of  the common  envelope efficiency  (0.2--0.3), a  result that  is in
agreement  with  previous  findings   obtained  by  observational  and
population synthesis  studies of close  SDSS WD+MS binaries.   We also
show  that all  synthetic  populations that  result  from adopting  an
initial mass ratio distribution with  a positive slope are excluded by
observations. Finally, we confirm that the properties of the simulated
WD+MS binary populations are nearly independent of the age adopted for
the thin  disk, on the contribution  of WD+MS binaries from  the thick
disk  (0--17 per  cent of  the total  population) and  on the  assumed
fraction of  the internal energy  that is  used to eject  the envelope
during the common envelope phase (0.1--0.5).

\end{abstract}

\begin{keywords}
(stars:) binaries (including multiple): close~--~(stars):
white dwarfs~--~(stars:) binaries: spectroscopic
\end{keywords}


\section{Introduction}

White  dwarf-main  sequence  (WD+MS)  binaries  are  the  evolutionary
products of main sequence binaries. In  $\sim$75 per cent of the cases
the initial main sequence binary  separations are large enough for the
binary components  to evolve in  the same way  as if they  were single
stars \citep{Willems2004}.   The orbital separations of  the remaining
$\sim$25 per cent  of main sequence binaries are close  enough for the
systems to undergo a phase  of dynamically unstable mass transfer once
the  primary  becomes  a  red  giant or  an  asymptotic  giant  branch
star. This  leads to  the formation  of a  common envelope  around the
nucleus  of   the  giant   star  and   the  main   sequence  companion
\citep{Webbink2008}, and hence  to a dramatic decrease  of the orbital
separation.   WD+MS binaries  that evolved  through a  common envelope
phase are known as post-common envelope binaries (PCEBs).

Modern  large scale  surveys  such  as the  Sloan  Digital Sky  Survey
\citep{York2000}, the  UKIRT Infrared  Sky Survey  \citep{Dye2006} and
the Large sky Area Multi-Object fiber Spectroscopic Telescope (LAMOST)
survey   \citep{Zhao2012},  have   facilitated   the  compilation   of
comprehensive spectroscopic  WD+MS binary samples during  the last few
years \citep{Heller2009,  RM2010,RM2012a,Liu2012,RM2013,Ren2014}.  The
SDSS has  been a particularly rich  source for the discovery  of WD+MS
binaries, mainly  due to  their overlap in  colour space  with quasars
\citep{Smolcic2004}. Roughly  25 per cent  of the SDSS  WD+MS binaries
are PCEBs  \citep{NM2011}, some  of which have  been identified  to be
eclipsing    \citep{NM2009,Pyrzas2009,Parsons2013,Parsons2015}.    The
largest  and most  homogeneous catalogue  of WD+MS  binaries currently
available is that from \cite{RM2016}, with a total number 3291 systems
identified within the data release 12 of SDSS.

Observationally,  SDSS WD+MS  binaries  have been  used  as tools  for
analyzing several  open and  interesting problems. These  include, for
example,  constraining  theories  of close  compact  binary  evolution
\citep{Zorotovic2010,Davis2010,DeMarco2011,Zorotovic2011,RM2012b},
providing  observational confirmation  for disrupted  magnetic braking
\citep{Schreiber2010,Zorotovic2016}, rendering robust evidence for the
majority   of  low-mass   white  dwarfs   being  formed   in  binaries
\citep{RM2011}, studying the pairing properties of main sequence stars
\citep{Ferrario2012}, using  WD+MS binaries as new  gravitational wave
verification sources  \citep{Kilic2014}, testing  the existence  of an
age-metallicity relation  in the Galactic disk  \citep{RM2016b}, using
SDSS eclipsing  PCEBs to  test the  existence of  circumbinary planets
\citep{Zorotovic2013,Marsh2014}.

Theoretically, population  synthesis studies reproducing  the ensemble
properties of SDSS WD+MS binaries  have been also highly successful at
improving  our  current  understanding of  common  envelope  evolution
\citep{Toonen2013,Camacho2014,Zorotovic2014}. However, it is important
to keep  in mind that  these works focused  mainly on the  PCEB sample
observed by the SDSS.  Clearly, far more can be learned from analyzing
the  total (PCEB  plus  wide  WD+MS binary)  population.  In order  to
overcome  this drawback,  in  this paper  we present  a  suite of  new
simulations of  the entire  population of WD+MS  binaries in  the SDSS
data release  12.  To this  end we employ  an improved version  of the
population synthesis  code used  in \citet{Camacho2014}.  Our  code is
based  in Monte  Carlo techniques  and  takes into  account all  known
observational  biases  and  selection   procedures.   Also,  we  adopt
different star formation histories,  initial mass ratio distributions,
common  envelope   efficiencies,  thin   disk  ages  and   a  variable
contribution of WD+MS  binaries from the thick disk. All  this aims at
constraining  which set  of  parameters fit  better the  observational
data.

The  paper  is organized  as  follows.   In Section~\ref{sec:data}  we
describe  the  observational  sample   of  SDSS  WD+MS  binaries.   In
Section~\ref{sec:code} we  describe our  Monte Carlo simulator  and in
Section~\ref{sec:obsv} we  explain how all known  observational biases
are     implemented    in     our    numerical     simulations.     In
Section~\ref{sec:results-4} we present our results. We close our paper
by  summarizing  our  main  results and  drawing  our  conclusions  in
Section~\ref{sec:conc}.


\section{The observed WD+MS sample}
\label{sec:data}

As  already  mentioned,  the  SDSS WD+MS  binary  catalogue  currently
constitutes the  largest and most homogeneous  sample of spectroscopic
WD+MS  binaries,  with  3,291   systems  identified  within  the  data
release 12  \citep{RM2016}.  Because  of selection  effects, the  vast
majority of SDSS  WD+MS binaries contain a  low-mass M-dwarf secondary
star, as hotter main sequence stars generally outshine the white dwarf
in the optical SDSS spectrum (more  details on this issue are provided
in Section\,\ref{sec:intrinsic}).

\begin{figure}
  \centering
  \includegraphics[width=\columnwidth]{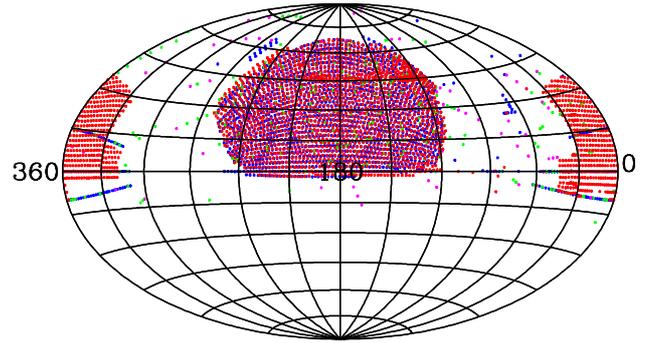}
  \caption{A Hammer-Aitoff projection in equatorial coordinates of the
    SDSS  spectroscopic  plate  positions. Red  dots  indicate  Legacy
    plates, blue  dots are  used for  BOSS, green  dots for  SEGUE and
    magenta dots for SEGUE-2}
  \label{fig:plates}
\end{figure}

The majority of SDSS WD+MS binaries  have been observed as part of the
Legacy  Survey  \citep{AM2008,Abazajian2009}  and  BOSS  \citep[Barion
Oscillation  Spectroscopic  Survey;][]{Dawson2013} simply  because  of
their overlap  in colour space  with quasars.  Hence, Legacy  and BOSS
WD+MS  binaries  generally  contain hot  ($\gtrsim  10,000\,$K)  white
dwarfs. In order  to overcome this observational  bias, WD+MS binaries
were additionally observed as part of a SEGUE --- Sloan Exploration of
Galactic Understanding  and Evolution \citep{Yanny2009}  --- dedicated
survey aimed at targeting WD+MS  binaries containing cool white dwarfs
\citep{RM2012a}.   Hereafter, we  flag  these systems  as SEGUE  WD+MS
binaries.  Finally, a small number  of WD+MS binaries were observed by
the SEGUE and SEGUE-2 surveys of  SDSS that aimed at obtaining spectra
of main sequence stars and red  giants. We flag these as SEGUE-2 WD+MS
binaries.

White dwarf  effective temperatures, surface gravities  and masses and
secondary star (M  dwarf) spectral types were derived for  each of the
SDSS   WD+MS   binaries   from    their   SDSS   spectra   using   the
decomposition/fitting          routine           described          in
\citet{RM2007}. \citet{RM2016} demonstrated that the stellar parameter
distributions  resulting from  the four  different sub-populations  of
SDSS  WD+MS binaries  (namely Legacy,  BOSS, SEGUE  and SEGUE-2  WD+MS
binaries) are statistically different. This is a simple consequence of
the different  selection criteria and  magnitude cuts employed  by the
four  different  sub-surveys.  Thus,  the  overall  SDSS WD+MS  binary
population  is   severely  affected   by  selection  effects.    As  a
consequence, modeling the entire SDSS WD+MS binary sample entails some
complications, as implementing the  specific selection biases for each
sub-sample is  needed.  Moreover,  our simulations  need to  take into
account that SDSS observed in specific regions of the sky, being these
marked  by the  positions of  the over  4,000 fibre-fed  spectroscopic
plates used during the observations (see Fig.~\ref{fig:plates}).


\section{The synthetic WD+MS sample}
\label{sec:code}

In this section  we describe our Monte Carlo simulator  and we provide
details on our adopted evolutionary  sequences and white dwarf cooling
tracks.

\subsection{The population synthesis code}

To simulate  the population of  WD+MS binaries we employed  an updated
version of  the population  synthesis code  designed to  simulate both
single and binary  white dwarfs in the Galactic  disk, first presented
in \cite{Camacho2014}  and built as  an extension of a  previous Monte
Carlo code  \citep{GB1999,GB2004,Torres2005} specifically  designed to
study the population  of single white dwarfs in the  Galactic disk. In
what follows we describe its main ingredients.

We start  providing the initial conditions  for any of the  stars that
constitute our population. These are the zero-age main sequence (ZAMS)
mass   sampled    from   the   initial   mass    function   (IMF)   of
\citet{Kroupa2001}, the time of birth obtained from the star formation
history (SFH)  --- which is  assumed to  be constant in  our reference
model  --- the  metallicity (assumed  to be  Solar), and  a single  or
binary star membership  according to an adopted binary  fraction of 50
per  cent \citep{Duchene2013}.   If the  star  is member  of a  binary
system we obtain  the ZAMS mass of the secondary  star according to an
initial mass ratio distribution (IMRD) --- which is assumed to be flat
in  our reference  model.  The  initial  separation of  the binary  is
chosen from a logarithmically  flat distribution \citep{Davis2010} and
the   initial  eccentricity   from  a   linear  thermal   distribution
\citep{Heggie1975}. We  then assign  a location (and,  consequently, a
distance) of the  object.  The equatorial coordinates for  each of our
synthetic star follow the distributions of  all SDSS plates used up to
the  data release  12 (Fig.~\ref{fig:plates})  adopting a  solid angle
aperture of 7  square degrees per plate. The radial  distance from the
Sun extends up to distances of 2~kpc. To distribute synthetic stars we
employ a double  exponential function for the local  density of stars,
with a scale  height of 300~pc and  a scale length of  2.6~kpc for the
thin disk, and a 900~pc scale  height and 3.5~kpc scale length for the
thick disk \citep{BH2016}. Our density distribution is then normalized
to  the  local mass  density  within  200~pc  from the  Sun,  adopting
standard values, namely  $\rho_{\rm thin}=0.094\, M_{\sun}/{\rm pc^3}$
for the thin  disk \citep{Holmberg2000} and thick  disk --- $\rho_{\rm
  thick}=8.5$  per cent  of  $\rho_{\rm  thin}$ \citep{Reid2005}.   We
finally  compute velocities  in  the  standard right-handed  Cartesian
Galactic system where the mean  space velocity components for the thin
disk population  are $(\langle U\rangle  ,\,\langle V\rangle\,,\langle
W\rangle\,)=(-8.62\,,-20.04\,,-7.10)\,$km/s   and  its   corresponding
dispersions            are            $(\sigma_U\,,\sigma_V,\sigma_W)=
(32.4\,,23.0\,,18.1)\,$km/s,  while  for  the   thick  disk  we  adopt
$(\langle        U\rangle         ,\,\langle        V\rangle\,,\langle
W\rangle\,)=(-11.0\,,-42.0\,,-12.0\,)\,$km/s                       and
$(\sigma_U,\sigma_V,\sigma_W)=  (50.0\,,54.0\,,34.0\,)\,$km/s ---  see
\cite{Rowell2011} for details.

We then  allow the synthetic single  or binary system to  evolve until
present time,  adopting for  our reference  model a  thin disk  age of
10~Gyr \citep{Cojocaru2014} and  a thick disk age of  12~Gyr.  This is
motivated by the findings of \cite{Felzing2009} who presented a sample
of very likely  thick disk candidates with ages on  average well above
10~Gyr  and of  \cite{Ak2013} who  found that  thick disk  cataclysmic
variables have ages up to 13~Gyr.  If the synthetic star is single and
has time  to become a  white dwarf,  it evolves following  the cooling
tracks detailed  in the following section.   If that is the  case, the
mass of  the white  dwarf is obtained  from the  initial-to-final mass
relation    (IFMR)     according    to    the     prescription    from
\cite{Hurley2002}. If the object is member  of a binary system and the
primary  star has  time to  become a  white dwarf,  then the  pair can
evolve through  two different  scenarios.  In  the first  scenario the
binary evolves without mass transfer interactions as a detached system
and  the primary  star evolves  into a  white dwarf  that subsequently
cools  down following  the  cooling sequences  described  in the  next
section.  In this case the mass  of the white dwarf is also calculated
from  the initial-to-final  mass relation  of \cite{Hurley2002}.   The
second scenario involves  mass transfer episodes and  the evolution of
the binary is obtained following  the prescriptions of the BSE package
\citep{Hurley2002},  following the  parameter assumptions  detailed in
\cite{Camacho2014}.  If the system  evolves though the common envelope
phase we  use the $\alpha$-formalism as  described in \cite{Tout1997},
with $\alpha_{\mathrm{CE}}$ being the efficiency in converting orbital
energy into kinetic energy to eject the envelope (assumed to be 0.3 in
our reference model). This implementation  also takes into account the
$\alpha_{\mathrm{int}}$ parameter (assumed to  be 0.0 in our reference
model), first presented in  \cite{Han1995}, describing the fraction of
the internal energy (thermal, radiation and recombination energy) used
to  eject  the  envelope.   As described  in  \cite{Camacho2014},  the
$\alpha_{\mathrm{int}}$ parameter  is used  to include the  effects of
the internal energy  in the binding energy  parameter $\lambda$, which
is  thus not  taken  as  a constant,  but  computed  using a  specific
algorithm \citep{Claeys2014}  in BSE.  In the  current version  of the
code,  provided  that   a  positive  value  is   used,  the  parameter
$\alpha_{\mathrm{int}}$  represents  the   fraction  of  recombination
energy that  contributes to  eject the envelope.   It is  important to
note that  the thermal  energy of  the envelope  is always  taken into
account (using the virial  theorem) even if $\alpha_{\mathrm{int}}$ is
set to zero. For a more detailed discussion on how this is implemented
in the latest version of BSE and important comments on the correct use
of BSE and the notations used in the code itself, we direct the reader
to    \cite{Zorotovic2014b},    mentioning    that    the    notations
$\alpha_{\mathrm{int}}$ or  $\alpha_{\mathrm{rec}}$ are, in  our case,
equivalent.

\subsection{Evolutionary sequences and cooling tracks}

The BSE package \citep{Hurley2002} provides luminosities, temperatures
and surface gravities for both the  main sequence and the white dwarf,
computed  using  the  evolutionary  tracks of  \cite{Pols1998}  and  a
modified Mestel cooling law, respectively. We re-compute these stellar
parameters  using more  modern  tracks that  also provide  photometric
magnitudes in the Johnson-Cousins  $UBVRI$ system (taking into account
both  rejuvenation  and  ageing   during  overflow  episodes).   These
evolutionary tracks are also used  to derive the stellar parameters of
the binary components in binaries  where no mass transfer interactions
take place.

For  carbon-oxygen white  dwarfs  (that is,  white  dwarf with  masses
$M_{\rm{WD}}$  between  $0.45$  and  $1.10\,  M_{\sun}$)  we  use  the
evolutionary calculations  of \cite{Renedo2010}, for  oxygen-neon core
white dwarfs  (namely, white dwarfs  with masses $M_{\rm{WD}}  > 1.1\,
M_{\sun}$)    we    employ    those    of    \cite{Althaus2005}    and
\cite{Althaus2007},  and for  helium  core white  dwarfs (with  masses
$M_{\rm{WD}}  <0.45\,  M_{\sun}$)  we   used  the  cooling  tracks  of
\cite{Serenelli2001}.    In   all    cases   Solar   metallicity   and
hydrogen-rich  atmospheres are  assumed and  the full  set of  $UBVRI$
magnitudes are provided.

For the main sequence companion  we employ the new evolutionary tracks
for  low  mass stars  of  \cite{Baraffe2015},  which provide  improved
predictions for optical colours. The  only downside here is that these
sequences only  provide the $VRI$  magnitudes, $U$ and  $B$ magnitudes
are recovered  using a  third order  polynomic approximation  based on
observations of G, K and M stars of \cite{Pickles1998}.  The effective
temperatures are converted into  spectral types following the spectral
type-effective temperature relation of \citet{Camacho2014}.

For both the main sequence stars and the white dwarfs we converted the
$UBVRI$   magnitudes  into   the   SDSS  $ugriz$   system  using   the
transformations described in  \cite{Jordi2006}. Reddening was computed
taking into  account the  position of  the star  using the  results of
\cite{Hakkila1997}     and     the     updated     coefficients     of
\cite{Schlafly2010}.


\section{Observational biases}
\label{sec:obsv}

In this  section we provide  details about the selection  effects that
affect the  observed SDSS WD+MS  binary population and we  explain how
these are incorporated into our synthetic population models.

\begin{figure}
  \centering
  \includegraphics[width=\columnwidth]{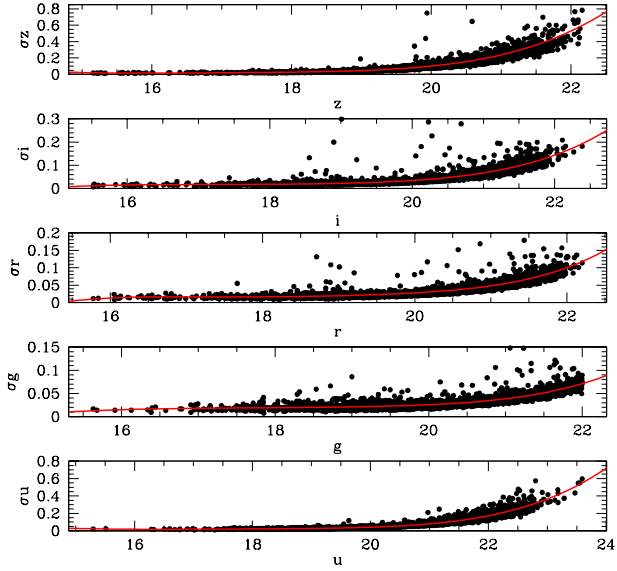}
  \caption{Distribution of  photometric errors  in the $u$,  $g$, $r$,
    $i$ and $z$  (black dots) passbands, each fitted to  a fifth order
    polynomial (red line).}
  \label{fig:error_mag}
\end{figure}

\subsection{Colour and magnitude cuts}

The  SDSS spectroscopic  survey is  magnitude-limited. Hence,  all our
synthetic populations of WD+MS binaries must comply with the magnitude
cuts of the SDSS, which are the following ones:

\begin{align}
15 < &\, i < 19.1 \textrm{, for Legacy} \\
15 < &\, g < 22 \textrm{, for BOSS} \\
15 < &\, g < 20 \textrm{, for SEGUE and SEGUE-2}
\end{align}

Moreover, observed  SDSS WD+MS binaries  define a clear region  in the
$ugriz$ colour space  \citep{RM2013} which allows us  to define colour
cuts to  cull our synthetic samples  (see Fig.~\ref{fig:col_diag2} for
an example):

\begin{align}
(u-g) < &\, 0.93 - 0.27 \times (g-r) - 4.7 \times (g-r)^2 + \nonumber \\
&\, 12.38 \times(g-r)^3 + 3.08 \times (g-r)^4 - 22.19 \times \nonumber \\
&\, (g-r)^5 + 16.67 \times(g-r)^6 - 3.89 \times (g-r)^7,
\end{align}
\begin{align}
\, & -0.6 <(u-g), \nonumber \\
(g-r) < &\,\, 2 \times (r-i) + 0.38 \textrm{,\,\,\,\,if } -0.4 < (r-i)\leq 0.06, \nonumber \\
(g-r) < &\,\, 4.5 \times (r-i) - 0.85 \textrm{,\,\,\,\,if } 0.3 < (r-i) \leq 
0.48, \nonumber \\
(g-r) < &\,\, 0.5 \textrm{,\,\,\,\,if } 0.06 < (r-i) \leq 0.3, \nonumber \\
\,& -0.5<(g-r)<1.3, \nonumber \\
\,& -0.4<(r-i)<1.6, \nonumber \\
(r-i) < &\,\, 0.5 + 2 \times (i-z) \textrm{,\,\,\,\,if } (i-z) > 0, \nonumber \\
(r-i) < &\,\, 0.5 + (i-z) \textrm{,\,\,\,\,if } (i-z) \leq 0, \nonumber \\
\, & -0.8<(i-z)< 1.15
\end{align}

It is important  to emphasize here that the SEGUE  WD+MS binary survey
was defined for targeting WD+MS binaries containing cool white dwarfs.
Hence, SEGUE WD+MS binaries populate  different regions in the $ugriz$
colour  space \citep{RM2012a}.   These  regions  define the  following
colour  cuts  that  we  apply  to the  synthetic  SEGUE  WD+MS  binary
population:

\begin{align}
\,&\,(u-g) < 2.25, \nonumber\\
\,-0.2 <\,&(g-r) < 1.2, \nonumber\\
\,&(g-r) > -19.78(r-i)+11.13, \nonumber\\
\,&(g-r) < 0.95\times (r-i)+0.5, \nonumber\\
\,&(i-z) > 0.5 \textrm{,\,\, for } (r-i) > 1.0, \nonumber\\
\,&(i-z) > 0.68\times (r-i)-0.18 \textrm{,\,\, for } (r-i) \leq 1.0, \nonumber\\
\,0.5<\,&(r-i) < 2.0, \nonumber\\
15\, <\,&g < 20
\end{align}

\begin{figure}
  \centering
  \includegraphics[width=\columnwidth]{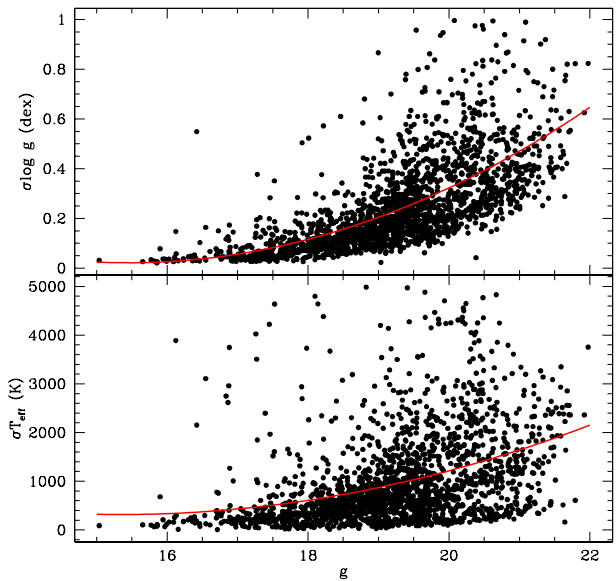}
  \caption{Errors for  white dwarf effective temperatures  and surface
    gravities  as a  function  of the  apparent  magnitude $g$  (black
    dots),  fitted  to  a  third order  polynomial  distribution  (red
    line).}
\label{fig:error_param}
\end{figure}

\subsection{Spectroscopic completeness}

The target selection criteria employed by the different sub-surveys of
the SDSS implies that not all WD+MS binaries have the same probability
of  being  observed. The  Legacy  and  BOSS surveys  follow  selection
criteria that aim at targeting mainly galaxies \citep{Strauss2002} and
quasars   \citep{Richards2002,  Ross2012}.    Hence,  WD+MS   binaries
containing  hot white  dwarfs ($>10,000-15,000$\,K)  and/or late  type
($>$M2)  companions  are more  likely  to  be observed.   SEGUE  WD+MS
binaries  are  dominated by  systems  containing  cooler white  dwarfs
and/or  early type  companions, a  simple consequence  of the  defined
target  selection  criteria  \citep{RM2012a}. Finally,  SEGUE-2  WD+MS
binaries have similar colours to  those of single main sequence stars,
i.e.   the  white dwarf  contributes  little  to the  spectral  energy
distribution. Hence, in order to  produce realistic simulations of the
SDSS WD+MS population we need to implement the probability for a given
WD+MS binary to be observed. That is, we need to apply a spectroscopic
completeness correction.

The  first  step in  this  process  is determining  the  spectroscopic
completeness of the observed  sample, following the approach described
in \cite{Camacho2014}. That  is, we consider a  four dimensional space
composed by the $u-g$, $g-r$, $r-i$ and $i-z$ colours, and we define a
$0.2$ magnitude four-dimension sphere  around each observed source. We
then use  the {\tt  casjobs} interface  to count  the number  of point
sources with clean  photometry ($N_{\rm phot}$) as well  as the number
of  spectroscopic sources  ($N_{\rm spec}$)  within each  sphere.  The
ratio $N_{\rm spec}/N_{\rm phot}$ gives the spectroscopic completeness
for the observed  system. Then, for each WD+MS binary  produced in the
synthetic  sample, we  compute a  four-dimensional distance  in colour
space  to each  of  the  observed WD+MS  binaries  and  we select  the
observed system that is closest to  the synthetic one. If the distance
to the selected closest observed binary is less than $0.2$ magnitudes,
then the synthetic WD+MS pair  will be assigned the same spectroscopic
completeness as that of the  observed one. Conversely, if the distance
is larger than  $0.2$ magnitudes, then the  assigned completeness will
be   null.    This   exercise   is  performed   separately   for   the
observed/simulated systems within the four sub-surveys of SDSS.

\subsection{Intrinsic WD+MS binary bias}
\label{sec:intrinsic}

In order  to detect  in the sample  a spectrum of  a WD+MS  binary the
spectral features of  both components must be  observed.  This implies
that  WD+MS binaries  in  which one  of the  two  stars dominates  the
spectral energy  distribution will be  harder, or even  impossible, to
detect \citep{Parsons2016}. Moreover, WD+MS  binaries that are further
away are intrinsically  fainter and the resulting SDSS  spectra are of
lower signal-to-noise  ratio (since  the SDSS exposures  are generally
the same for  all targets).  Identifying the spectral  features of the
two  components is  obviously  more difficult  when  dealing with  low
signal-to-noise ratio  spectra.  It is  then mandatory to  eliminate a
certain percentage of the synthetic  WD+MS binaries according to these
reasonings.

\cite{Camacho2014} presented a multi-dimensional  grid of WD+MS binary
parameters (white dwarf effective  temperatures and surface gravities,
secondary star spectral types and  distances) that allowed to evaluate
which  synthetic  WD+MS  binaries  would have  been  detected  by  the
SDSS. In this work we follow the same approach.

\begin{table}
\caption {Percentage  of WD+MS binaries  respect to the  entire sample
  that   survive  the   observational   biases   implemented  in   the
  simulations.  The values are provided for our reference model.}
\label{tab:percent}
\begin{center}
\begin{tabular}{lcc}
\hline
Observational bias  & Filtered (\%) & Cumulative (\%) \\ 
\hline
Colour and magnitude cuts       & $28$   & $28$ \\
Spectroscopic completeness      & $5$    & $1.4$\\
Intrinsic WD+MS binary bias     & $72$   & $1.0$\\
\hline
\end{tabular}
\end{center}
\end{table}

\subsection{Uncertainties in the observed parameters}

The measured  SDSS photometric  magnitudes and the  stellar parameters
derived from fitting the SDSS WD+MS binary spectra can have relatively
large  uncertainties  \citep{RM2016}.   Hence,   it  is  necessary  to
incorporate such  uncertainties in  the synthetic WD+MS  binary sample
before any  comparison to  the observational  data sets  is performed.
Fig.~\ref{fig:error_mag} shows the photometric  errors, $\sigma$, as a
function  of  the  corresponding  magnitude.   As  can  be  seen,  the
photometric  errors  clearly increase  as  the  apparent magnitude  is
fainter. We  fitted the  distributions using fifth  order polynomials,
that provide us  with an expression for $\sigma$ as  a function of the
apparent magnitude. We  then define a Gaussian  error distribution for
that  specific  magnitude  that  we  sample in  order  to  obtain  the
photometric error of each synthetic  WD+MS binary in each passband. We
apply a similar procedure for the  errors in the white dwarf effective
temperature and surface gravity, using a third order polynomial fit in
this case (see Fig.~\ref{fig:error_param}). For the companion spectral
type distribution we assumed a constant  value of $\sigma$ of one bin,
i.e.  an uncertainty of one  spectral sub-class. Only after adding the
corresponding errors in photometric  magnitudes and stellar parameters
we do apply the colour and  magnitude cuts and the other observational
filters  previously  described. Given  the  random  character of  this
procedure,  for each  realization  providing us  with  a WD+MS  binary
sample from  the Monte Carlo code,  we repeated the process  of adding
errors and afterwards filtering the sample 20 times per realization.

\begin{table}
\caption {Percentage of present day WD+MS binaries that have undergone
  common envelope evolution (and have not yet merged), in the complete
  and filtered  sample for different values  of $\alpha_{\mathrm{CE}}$
  and  $\alpha_{\mathrm{int}}$ assumptions.  Note that  the effect  of
  $\alpha_{\mathrm{int}}$   over    the   filtered    sample   becomes
  negligible.}
\label{tab:alphaCE}
\begin{center}
\begin{tabular}{cccc}
\hline
\hline
$\alpha_{\mathrm{CE}}$ & $\alpha_{\mathrm{int}}$ & Complete sample (\%) & 
Filtered sample (\%) \\ 
\hline
$0.1$ & $0.0$ & $3$ & $10$ \\
$0.2$ & $0.0$ & $6$ & $21$ \\
$0.3$ & $0.0$ & $10$ & $29$ \\
$0.3$ & $0.1$ & $13$ & $30$ \\
$0.3$ & $0.2$ & $14$ & $32$ \\
$0.3$ & $0.5$ & $17$ & $33$ \\
$0.5$ & $0.0$ & $16$ & $38$ \\
$0.7$ & $0.0$ & $21$ & $43$ \\
$1.0$ & $0.0$ & $26$ & $45$ \\
\hline
\hline
\end{tabular}
\end{center}
\end{table}

Finally,  for  both  the  observed   and  synthetic  samples  we  only
considered systems with  a relative error smaller than 10  per cent in
effective temperature, and  with absolute errors below  $0.075$ dex in
surface gravity.   This explains why different  distributions from the
same sample contain different numbers of WD+MS pairs.


\section{Results}
\label{sec:results-4}

We  used our  population  synthesis  code to  model  the WD+MS  binary
population in the Galactic disk. For this, we first defined a standard
model that  uses a  flat IMRD ($n(q)$),  an age for  the thin  disk of
$10$\,Gyr, a constant star formation rate, $\alpha_{\mathrm{CE}}=0.3$,
$\alpha_{\mathrm{int}}=0$  and  all  the fixed  parameter  assumptions
previously explained in Section~\ref{sec:code}.

\subsection{Preliminary checks}
\label{sec:prelim}

\begin{figure*}
  \centering 
  \includegraphics[width=1.5\columnwidth]{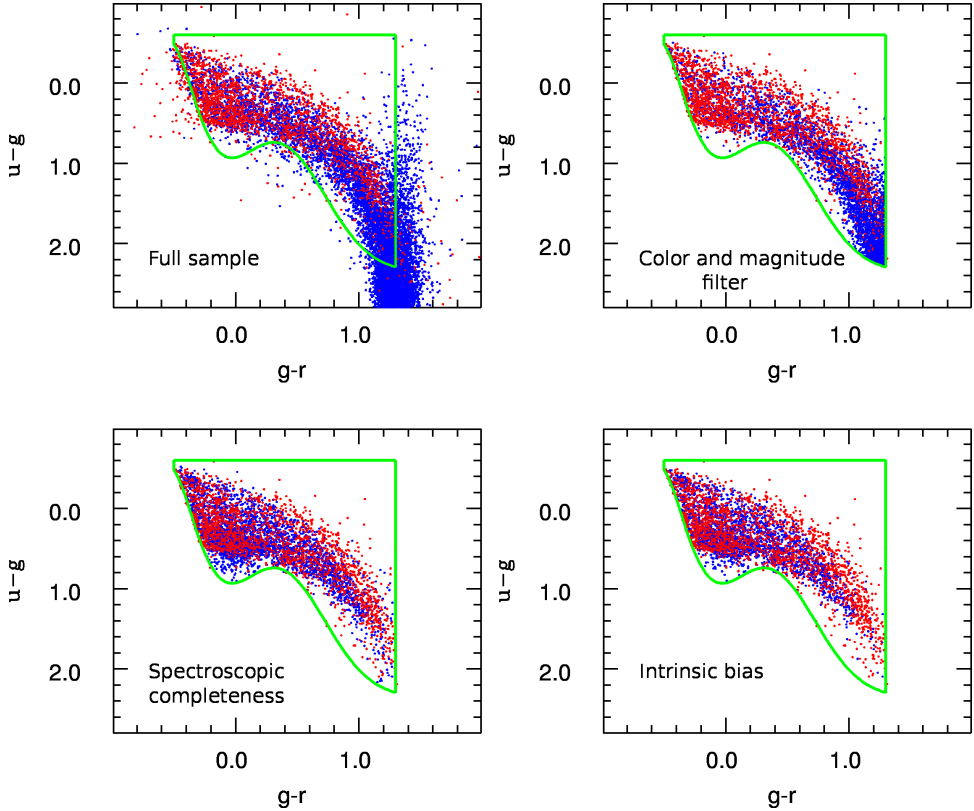}
  \caption{$u-g$ versus  $g-r$ colour-colour  diagram of  the observed
    (blue dots) and  synthetic (red dots) SDSS WD+MS  binaries for our
    reference model,  after incorporating the  different observational
    biases explained in Sect.~\ref{sec:obsv}.  See the on-line version
    of the journal for a colour version of this figure.}
  \label{fig:col_diag2}
\end{figure*}

First of all, we analyzed the effects of the observational biases.  In
Fig.\,\ref{fig:col_diag2} we  show a $u-g$ versus  $g-r$ colour-colour
diagram for  the synthetic SDSS  WD+MS binary population  that results
from our reference model (blue dots) compared with the observed sample
(red  dots). Each  panel  represents the  objects  that survive  after
consecutively applying the filters indicated  on it. Starting from the
full  sample (upper-left  panel), we  first  show the  effects of  the
colour and magnitude  filters (upper-right panel), the  effects of the
spectroscopic completeness  bias (lower-left panel) and,  finally, the
effect of applying the filter  for intrinsic bias (lower-right panel).
As can be seen in  Fig.\,\ref{fig:col_diag2} the number of binaries in
the initial synthetic sample decreases dramatically after applying the
different colour and magnitude cuts and observational biases. Also, we
notice that the final synthetic sample nicely matches the space colour
of  the observed  sample.  In  order  to provide  a more  quantitative
estimate   of    the   effect   of   the    selection   criteria,   in
Table\,\ref{tab:percent}  we show  the  percentage  of WD+MS  binaries
respect to  the full  sample that survive  these filters.   The second
column  represents the  percentage with  respect the  previous applied
filter and the third column  is the cumulative percentage with respect
the initial entire sample.  Inspection of the Table\,\ref{tab:percent}
reveals that  only 1  per cent  of the  entire simulated  WD+MS binary
population  makes it  to  the final  sample,  being the  spectroscopic
completeness the  most restrictive in percentage  of the observational
biases. These results  are in agreement with the analysis  of the PCEB
sample of \citet{Camacho2014}.

In order  to properly cover  the parameter space, we  initially varied
several input parameters to better understand their possible effect on
the  three distributions  under  scrutiny: the  white dwarf  effective
temperature and  surface gravity,  and the M  dwarf spectral  type. We
first varied  the age of the  thin disk between $8$  and $12$\,Gyr and
realised that the three  distributions were not particularly sensitive
to this  value.  We also  tried three different prescriptions  for the
star formation history: a constant rate (used in our reference model),
a  recent enhanced  star  formation with  one broad  peak  in the  SFR
between 1  and 3~Gyr ago  \citep{Vergely2002}, and a bimodal  SFR with
two broad peaks  at around 2 and 7~Gyr  ago \citep{Rowell2013}.  After
testing  each of  these  models independently,  we  conclude that  the
choice of the model of star formation has only marginal effects on the
synthetic $\log g$ distribution and  no other noticeable effect on the
other distributions that we analyze.

For   convenience,  in   our   simulations  we   used   the  IFMR   of
\cite{Hurley2002},   which  results   from   an  evolution   algorithm
consisting  basically of  a competition  between core-mass  growth and
envelope mass-loss.   Nevertheless, we tested this  procedure and used
the more reliable  IFMR of \cite{Catalan2008}. We  obtained that there
is a systematic  difference, although the difference of  masses of the
synthetic white dwarfs is generally not larger than $0.02\, M_{\sun}$.

Lastly, we  considered an initial  15 per cent contamination  of thick
disk stars, which  corresponds to a 7-17 per cent  contribution to the
final sample  (after all observational filters  are passed), depending
on the model  parameters. To quantify the effects  of the contribution
of thick disk systems we ran  one simulation in which the contribution
of the  thick disk was  suppressed and  we found that  the differences
distributions were minimal.

\subsection{The CE efficiency parameter}
\label{sec:res-CE}

The observational  analysis of  \citet{NM2011} showed that  between 21
and 24  per cent of SDSS  WD+MS binaries have experienced  a CE phase.
In Table~\ref{tab:alphaCE}  we provide,  for our reference  model, the
percentage of present-day  WD+MS binaries that have  evolved through a
CE episode (and did not become semi-detached nor merged) as a function
of  $\alpha_{\mathrm{CE}}$.   We did  this  for  the entire  simulated
sample  and for  the restricted  (filtered) sample  that survives  all
observational biases outlined  in Section\,\ref{sec:obsv}.  Inspection
of Table~\ref{tab:alphaCE} reveals that  the percentage of present-day
WD+MS  binaries that  underwent  a  CE phase  is  compatible with  the
observational        results       of        \citet{NM2011}       when
$\alpha_{\mathrm{CE}}=0.2-0.3$. For the sake  of completeness, we also
calculated    the   percentages    assuming   different    values   of
$\alpha_{\mathrm{int}}$  (ranging from  0.1 to  0.5).  We  found that,
although we obtain the same $\sim40$ per cent increase in PCEB systems
reported by \cite{Zorotovic2014} for  the complete sample, the results
are    nearly     independent    of     the    adopted     value    of
$\alpha_{\mathrm{int}}$, once  the observational filters  are applied.
Another interesting point would  be that observational biases actually
favor the discovery of PCEB systems.

\begin{figure}
  \centering
  \includegraphics[width=0.9\columnwidth]{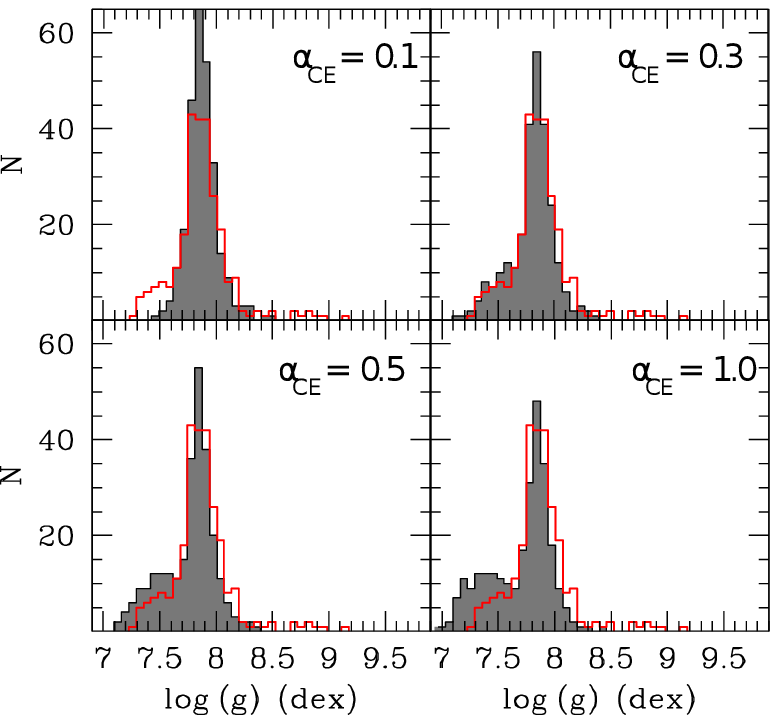}
  \caption{Distribution of  white dwarf $\log g$  for different values
    of $\alpha_{\mathrm{CE}}$ (with $\alpha_{\mathrm{int}}=0$)}
  \label{fig:logg}
\end{figure}

An additional  way to  constrain $\alpha_{\mathrm{CE}}$ is  to compare
the overall white  dwarf $\log g$ distribution of  the simulated WD+MS
binary   population  to   the   observed  one.    This   is  done   in
Fig.~\ref{fig:logg}.  By inspecting this  figure it becomes clear that
$\alpha_{\mathrm{CE}}=0.3$  reproduces  best the  observational  data.
This  result  is  in  agreement with  previous  observational  studies
\citep{Zorotovic2010} --- who found $\alpha_{\mathrm{CE}}=0.2-0.3$ ---
population   synthesis   studies   of  close   SDSS   WD+MS   binaries
\citep{Toonen2013,Camacho2014, Zorotovic2014}  --- who found  that the
CE  efficiency should  be low  --- and  even with  earlier theoretical
predictions  \citep{Taam2006} ---  who found  $\alpha_{\mathrm{CE}}\le
0.4-0.5$.

\begin{table*}
\centering
\caption{Our 12 models adopted for  the IMRD $n(q)$, ordered from best
  to worst according to the  fit to the observational $T_\mathrm{eff}$
  distributions, the value  of the three distance metrics  used in our
  analysis being averaged  between the results for  the Legacy survey,
  for the BOSS  survey and for the overall distribution  (see text for
  details).}
\label{tab:models-temp}
\setlength{\tabcolsep}{4ex}
\begin{tabular}{cccccc}
\hline
Model & $\langle D_{\rm KL}\rangle$ & Model &  $\langle D_{\rm LS}\rangle$ & Model & $\langle{\beta}\rangle\,(^{\circ})$\\
\hline
$3$  & $4.8\times 10^{-2}$ & $3$  & $4.9\times 10^{-3}$ & $3$  & $8.7$ \\
$4$  & $4.8\times 10^{-2}$ & $4$  & $6.0\times 10^{-3}$ & $6$  & $9.2$ \\
$6$  & $5.0\times 10^{-2}$ & $6$  & $6.0\times 10^{-3}$ & $4$  & $10.0$\\
$5$  & $5.4\times 10^{-2}$ & $1$  & $6.1\times 10^{-3}$ & $5$  & $10.2$\\
$1$  & $5.8\times 10^{-2}$ & $5$  & $6.3\times 10^{-3}$ & $1$  & $10.8$\\
$7$  & $6.5\times 10^{-2}$ & $7$  & $8.1\times 10^{-3}$ & $7$  & $12.2$\\
$2$  & $7.7\times 10^{-2}$ & $12$ & $1.5\times 10^{-2}$ & $12$ & $14.3$\\
$12$ & $7.9\times 10^{-2}$ & $2$  & $1.5\times 10^{-2}$ & $2$  & $14.5$\\
$8$  & $1.1\times 10^{-1}$ & $8$  & $2.5\times 10^{-2}$ & $8$  & $18.1$\\
$9$  & $1.2\times 10^{-1}$ & $9$  & $2.7\times 10^{-2}$ & $9$  & $18.8$\\
$10$ & $1.6\times 10^{-1}$ & $10$ & $3.6\times 10^{-2}$ & $10$ & $21.8$\\
$11$ & $2.4\times 10^{-1}$ & $11$ & $4.0\times 10^{-2}$ & $11$ & $23.1$\\
\hline
\end{tabular}
\end{table*}

\begin{table*}
\centering
\caption{Same as  Table \ref{tab:models-temp}, but using  the $\log g$
  distributions.}
\label{tab:models-logg}
\setlength{\tabcolsep}{4ex}
\begin{tabular}{cccccc}
\hline
Model & $\langle D_{\rm KL}\rangle$ & Model & $\langle D_{\rm LS}\rangle$ & Model & $\langle{\beta}\rangle\,(^{\circ})$\\
\hline
$5$  & $2.4\times 10^{-1}$ & $4$  & $9.0\times 10^{-3}$ & $6$  & $17.9$\\
$3$  & $2.4\times 10^{-1}$ & $3$  & $9.0\times 10^{-3}$ & $5$  & $18.1$\\
$6$  & $2.5\times 10^{-1}$ & $5$  & $1.0\times 10^{-2}$ & $4$  & $18.4$\\
$4$  & $2.6\times 10^{-1}$ & $1$  & $1.2\times 10^{-2}$ & $3$  & $18.6$\\
$1$  & $2.7\times 10^{-1}$ & $6$  & $1.2\times 10^{-2}$ & $1$  & $18.7$\\
$7$  & $2.8\times 10^{-1}$ & $7$  & $1.4\times 10^{-2}$ & $7$  & $19.3$\\
$8$  & $3.3\times 10^{-1}$ & $12$ & $1.7\times 10^{-2}$ & $8$  & $21.7$\\
$2$  & $4.4\times 10^{-1}$ & $2$  & $1.8\times 10^{-2}$ & $2$  & $21.7$\\
$12$ & $4.8\times 10^{-1}$ & $9$  & $2.0\times 10^{-2}$ & $12$ & $23.0$\\
$11$ & $6.3\times 10^{-1}$ & $8$  & $2.2\times 10^{-2}$ & $9$  & $24.7$\\
$9$  & $6.4\times 10^{-1}$ & $11$ & $3.8\times 10^{-2}$ & $11$ & $27.0$\\
$10$ & $1.1\times 10^{0}$  & $10$ & $4.2\times 10^{-2}$ & $10$ & $32.6$\\
\hline
\end{tabular}
\end{table*}

\begin{table*}
\centering
\caption{Same as  Table \ref{tab:models-temp},  but using the  M dwarf
  spectral  type distributions.  The first  five models  fit best  the
  observational data, the last seven models can be excluded. The seven
  IMRD dstributions that can be excluded all increase with $q$.}
\label{tab:models-spec}
\setlength{\tabcolsep}{4ex}
\begin{tabular}{cccccc}
\hline
Model & $\langle D_{\rm KL}\rangle$ & Model & $\langle D_{\rm LS}\rangle$ & Model & $\langle{\beta}\rangle\,(^{\circ})$\\
\hline
$6$  & $2.2\times 10^{-2}$ & $6$  & $8.3\times 10^{-3}$ & $6$  & $8.3$ \\
$3$  & $2.3\times 10^{-2}$ & $3$  & $1.1\times 10^{-2}$ & $3$  & $9.2$ \\
$5$  & $4.8\times 10^{-2}$ & $5$  & $2.0\times 10^{-2}$ & $5$  & $13.2$\\
$4$  & $5.0\times 10^{-2}$ & $4$  & $2.1\times 10^{-2}$ & $4$  & $13.5$\\
$1$  & $5.1\times 10^{-2}$ & $1$  & $2.2\times 10^{-2}$ & $1$  & $13.5$\\
\hline
$7$  & $8.1\times 10^{-2}$ & $7$  & $3.1\times 10^{-2}$ & $7$  & $16.8$\\
$2$  & $2.0\times 10^{-1}$ & $2$  & $5.7\times 10^{-2}$ & $2$  & $22.7$\\
$12$ & $2.8\times 10^{-1}$ & $12$ & $6.3\times 10^{-2}$ & $12$ & $24.7$\\
$10$ & $8.2\times 10^{-1}$ & $10$ & $1.1\times 10^{-1}$ & $10$ & $35.2$\\
$8$  & $9.3\times 10^{-1}$ & $8$  & $1.3\times 10^{-1}$ & $8$  & $37.6$\\
$9$  & $1.1\times 10^{0}$  & $9$  & $1.6\times 10^{-1}$ & $9$  & $39.4$\\
$10$ & $1.2\times 10^{0}$  & $11$ & $1.8\times 10^{-1}$ & $11$ & $41.7$\\
\hline
\end{tabular}
\end{table*}

\subsection{The initial mass ratio distribution}
\label{sec:imrd}

Our Monte Carlo simulator, calibrated using the largest sample of SDSS
WD+MS binaries currently known  \citep{RM2016} and taking into account
all known observational biases, offers  us an excellent opportunity to
constrain the properties of the IMRD, $n(q)$.  Hereafter we define the
mass ratio as $q=m_2/m_1$, where $m_1$  is the mass of the primary (or
more massive) star in a main sequence binary, and $m_2$ is the mass of
its  main  sequence companion  or  secondary  star.   The IMRD  is  an
important  tool for  understanding the  evolution of  stars in  binary
systems and  for constraining  models of  binary star  formation.  The
precise shape  of the IMRD  has been a topic  of much debate  for over
four decades,  with results  often contradicting each  other.  Indeed,
decreasing   \citep{Jaschek1972},  increasing   \citep{Dabbowski1977},
bimodal \citep{Trimble1974}  and flat \citep{Raghavan2010}  IMRDs have
been suggested.   This lack of  agreement remains when  recent results
are considered.  For example,  whilst \cite{Ducati2011} claim that the
IMRD decreases as $q$  increases, \cite{Reggiani2013} suggest a slowly
increasing IMRD.  These  discrepancies may be a  simple consequence of
the IMRD  being dependent  on both  the primary  mass and  the orbital
separation,  as suggested  by \citet{Duchene2013}.   This seems  to be
confirmed   by  several   observational  studies   \citep{Carrier2002,
  Burgasser2006,    Delfosse2004,     Carquillat2007,    Raghavan2010,
  Tokovinin2011, Sana2012, Reggiani2013,  Gullikson2016}.  However, it
is  important  to  emphasize  that  these  observational  studies  are
affected  by  important  selection  effects,  which  likely  introduce
sizable uncertainties  in the results. This  is particularly important
when the  secondary star  in a main  sequence binary  is intrinsically
faint and thus harder to detect against a moderately hot primary.

\begin{figure}
  \centering
  \includegraphics[width=0.9\columnwidth]{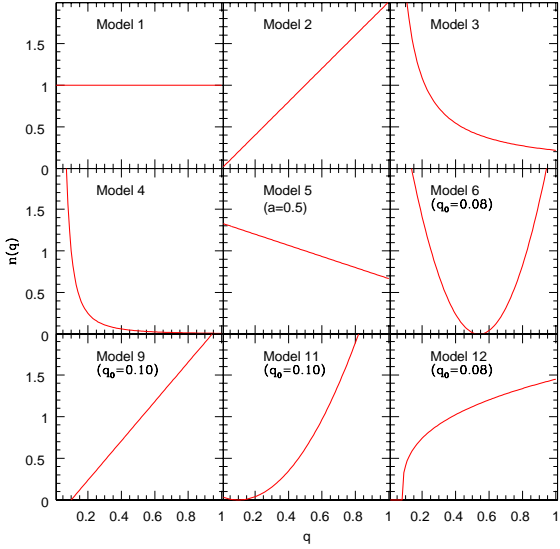}
  \caption{The 12 IMRD adopted in  our simulations. Note that the only
    difference between  models 6--7, 8--9  and 10--11 are  the assumed
    values of  $q_0$ ($0.08$ or $0.1$),  hence we only display  one of
    them. The distributions have been normalized to unity.}
  \label{fig:imrd}
\end{figure}

\begin{figure*}
  \centering
  \includegraphics[trim = 5mm 0mm 0mm 0mm, clip=true,width=1.2\columnwidth]{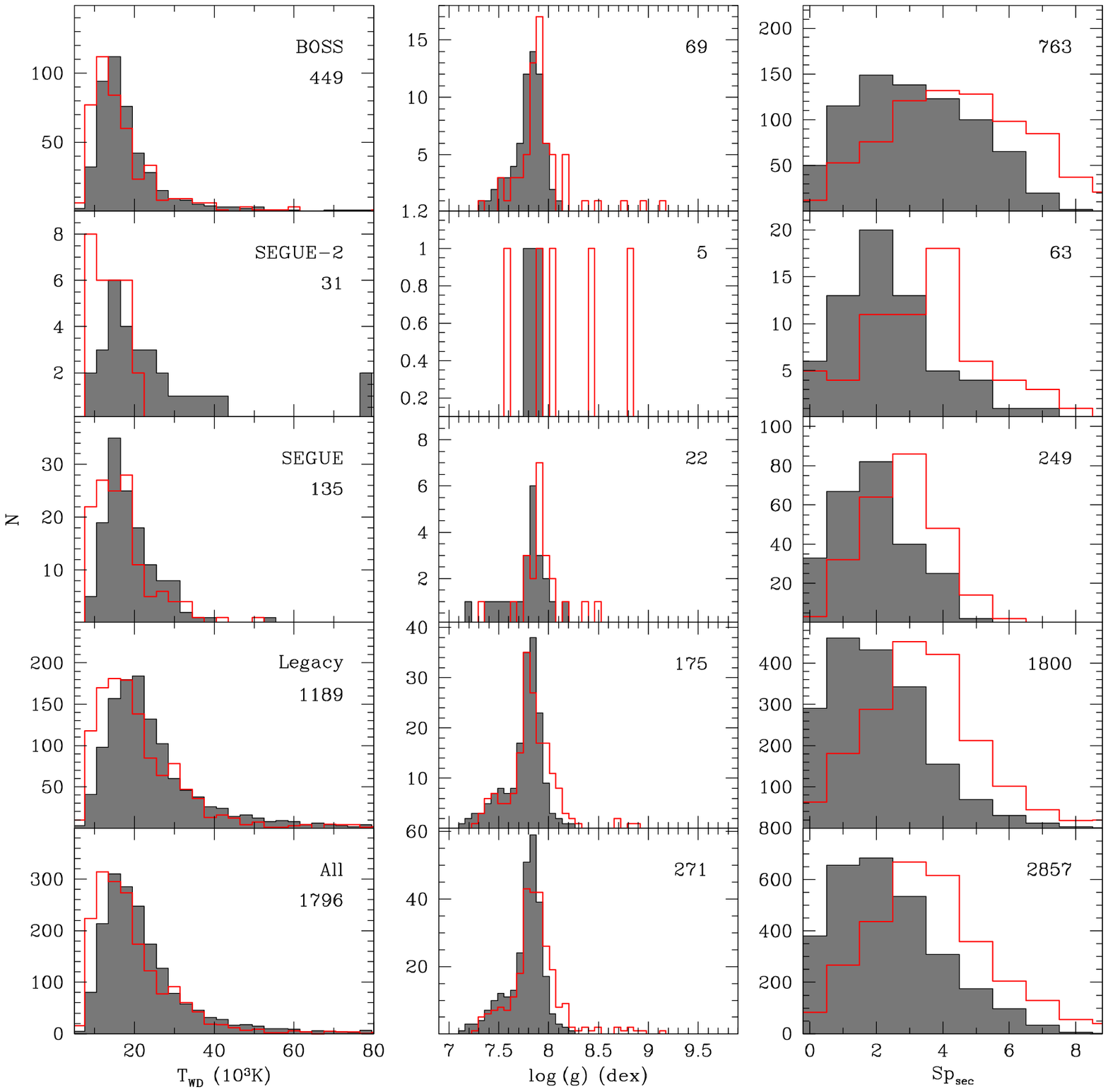}
  \caption{White dwarf  effective temperature,  surface gravity  and M
    dwarf  spectral  type   distributions.   Observational  data  from
    \citet{RM2016} (gray histograms), synthetic data obtained assuming
    a $n(q)  \sim q$  (red open histograms).  These panels  display an
    example of a bad fit.}
    \label{fig:model2}
\end{figure*}

\begin{figure*}
  \centering
  \includegraphics[trim = 5mm 0mm 0mm 0mm, clip=true,width=1.2\columnwidth]{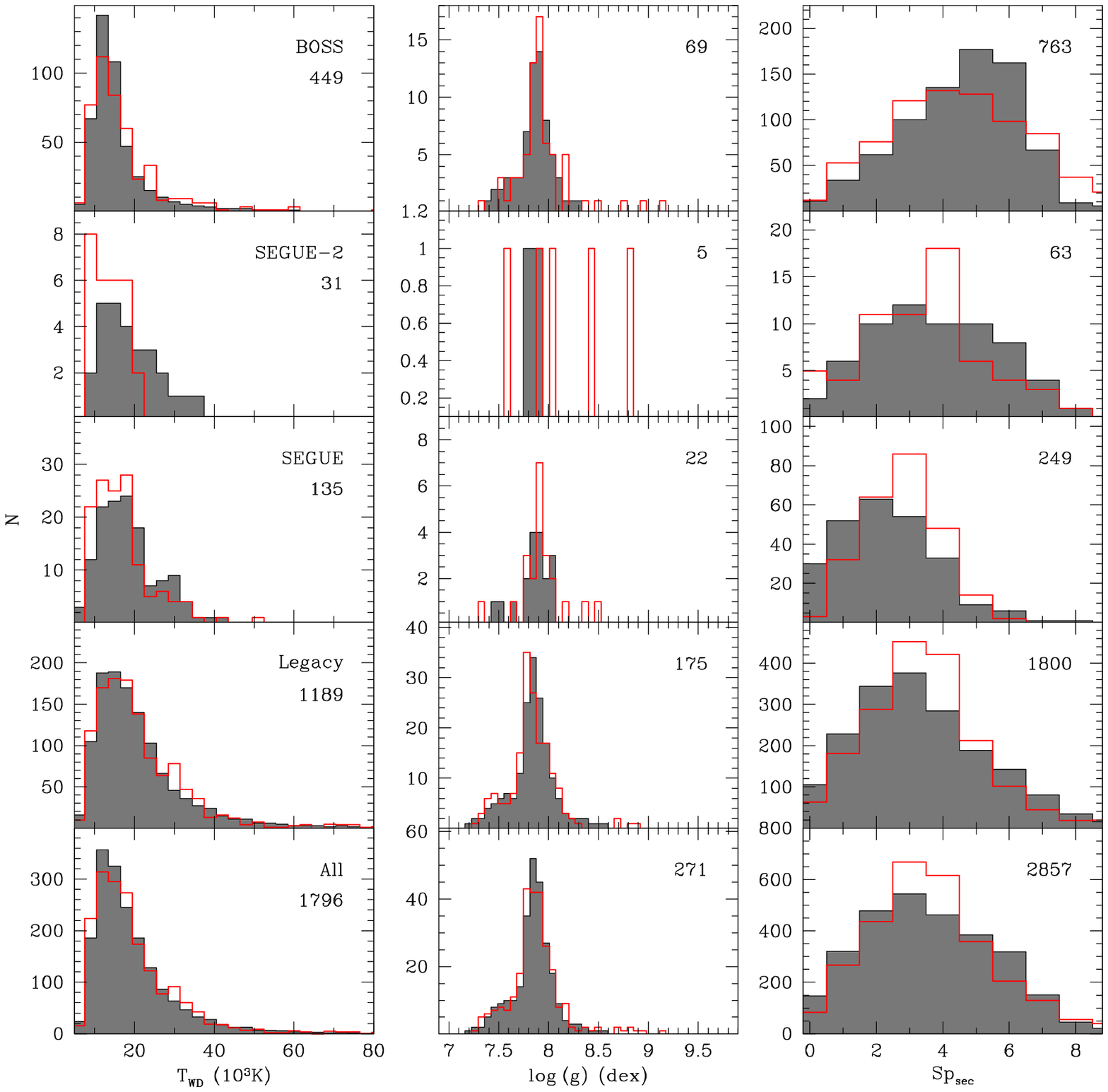}
  \caption{White dwarf  effective temperature,  surface gravity  and M
    dwarf  spectral   type  distributions.  Observational   data  from
    \citet{RM2016} (gray histograms), synthetic data obtained assuming
    a $n(q) \sim  q^{-1}$ (red open histograms). These  panels show an
    example of a good fit.}
    \label{fig:model3}
\end{figure*}

Recently, \cite{Ducati2011}  compared the  results of  a set  of Monte
Carlo simulations  of a  large number of  models with  different IMRDs
with a  sample of  249 binaries  belonging to  the Ninth  Catalogue of
Spectroscopic Binaries  \citep{Pourbaix2004}.  Their findings  favor a
linearly decreasing $n(q)  \sim 1-a q$, with  $a=0.5$.  However, given
the  diversity of  the  models they  studied, we  adopt  most of  them
together with  the more classical IMRD  studied in \cite{Camacho2014},
testing in total $12$ models for the IMRD.  We first adopted the three
most  frequently  assumed  IMRDs.   In  particular,  our  first  model
corresponds to $n(q) \propto 1$ (model 1).  Hence for model 1 we use a
flat  distribution, which  will be  regarded as  the reference  model.
Next we also used two  frequently employed IMRDs, namely $n(q) \propto
q$ (model 2) and $n(q) \propto q^{-1}$ (model 3). The other six models
that we used are less frequently employed, and are the following ones.
Model     4      corresponds     to     $n(q)      \propto     q^{-2}$
\citep{Hogeveen1992}. Model  5 is the best  model of \cite{Ducati2011}
$n(q) \propto 1-a \, q$, being $a=0.5$.  For models 6 and 7 we adopted
$n(q) \propto \left(q-\left(1+q_0\right)/2 \right)^2$, with $q_0=0.08$
and $q_0=0.10$, respectively. These models are inspired in the bimodal
distributions of \cite{Trimble1974}.  In models  8 and 9 we used $n(q)
\propto q-q_0$ with  $q_0=0.08$ and $q_0=0.10$, whereas  for models 10
and 11  the dependence on  $q$ is steeper $n(q)  \propto (q-q_0)^{2}$,
with $q_0=0.08$  and $q_0=0.10$,  respectively.  These models  peak at
$q=1.0$, as in the data  of \citealt{Fisher2005}. Finally, the IMRD of
model  12  is based  on  the  findings of  \cite{Reggiani2013},  $n(q)
\propto (q-q_0)^{1/3}$,  with $q_0=0.08$.  In  Fig.\,\ref{fig:imrd} we
display the 12 IMRDs.

The best  approach for  comparing the  simulated distributions  to the
observed ones is to use distance metrics, a procedure which will allow
us  not only  to decide  which  simulated distribution  fits best  the
observed data, but also to order  the models from best to worst. Among
the possible number of distance metrics  that can be defined, we chose
three commonly employed ones. Let $P$ be the observed distribution and
$Q$  the  simulated one,  the  three  metrics  employed here  are  the
following ones. We first employed a standard least squares method:
\begin{equation}
D_{\rm LS} = \sum_{i} (P(i)-Q(i))^2
\end{equation}
We also used the so-called Kullback-Leibler (KL) divergence:
\begin{equation}
D_{\rm KL} = \sum_{i} P(i) \ln\left(\frac{P(i)}{Q(i)}\right)
\end{equation}
Finally, we employed a less  known method, the so-called Bhattacharyya
coefficient:
\begin{equation}
\cos(\beta) = \sum_{i} \sqrt{P(i) Q(i)}
\end{equation}
The least squares method is a  standard distance metric.  On the other
hand, the Kullback-Leibler divergence is not symmetrical and both this
distance metric and  the Bhattacharyya coefficient do  not satisfy the
triangle  inequality, thus  they must  be considered  pre-metrics. The
Bhattacharyya   coefficient  also   has  an   interesting  geometrical
interpretation,   as   the   cosine   of   the   angle   between   two
multidimensional vectors describing the two distributions.

We employed these three methods and  ordered the models from lowest to
largest distance  (or angle), using  the $T_{\rm eff}$ and  $\log\, g$
distributions for the  white dwarf and the  spectral type distribution
of  the M  dwarf,  from Legacy  and  BOSS data  and  also the  overall
distributions.   We did  not use  SEGUE and  SEGUE-2 data  because the
sample sizes  are small. The overall  results are shown in  an ordered
way   in   Tables~\ref{tab:models-temp},   \ref{tab:models-logg}   and
\ref{tab:models-spec}.

Inspection of  Tables~\ref{tab:models-temp}, \ref{tab:models-logg} and
\ref{tab:models-spec} reveals  that we  roughly obtain the  same order
(from best  to worst model)  independently of  which one of  the three
metrics is used.   These types of distance metrics  are generally used
in optimizations  and it  is well  known that the  order they  give is
reliable. However,  they cannot  be used to  assess which  model truly
offers a  good fit ---  that is,  which models reproduce  the observed
distribution --- and which do not.  In order to further understand and
quantify this issue we performed  the following test.  We computed the
angle  which   defines  the  Bhattacharyya  coefficient   between  two
distributions  sampled from  the  same  model ---  that  is, the  same
parameter set  --- using  a different initial  random seed.   We found
that  typically we  obtain  a value  of $\beta$  ranging  from $7$  to
$10^{\circ}$  (on  average  $8.5^{\circ}$) for  the  distributions  of
$T_\mathrm{eff}$,  between  $11$  and  $17^{\circ}$ (with  a  mean  of
$14^{\circ}$) for the  distributions of $\log g$, and  between $3$ and
$6^{\circ}$ (with an  average value of $4.5^{\circ}$)  for the M-dwarf
spectral type distribution.

Keeping the  average values of $\beta$  in mind and comparing  them to
the    angles    listed     in    Tables\,\ref{tab:models-temp}    and
\ref{tab:models-logg}   for   the   $T_{\rm   eff}$   and   $\log   g$
distributions,  it can  be easily  realized that  for most  models the
angles between the theoretical distributions and the observed data are
smaller   than   2  times   the   average   values.   Conversely,   in
Table~\ref{tab:models-spec},  which  gives   the  angles  between  the
M-dwarf spectral type  distributions, it can be seen that  only two of
the 12 models are within an angle smaller than 2 times the mean angle.
This clearly indicates that the  M dwarf spectral type distribution is
much  more  sensitive to  the  choice  of the  IMRD.   We  thus set  a
threshold of 3  times the average angle for the  M-dwarf spectral type
distributions (i.e.  $13.5^{\circ}$) below which we consider the IMRDs
to be representative of the  observed data.  Interestingly, under this
assumption, we can exclude the seven IMRDs with increasing slopes (see
Table~\ref{tab:models-spec}  and Fig.\,\ref{fig:imrd}).   In order  to
exemplify    this,    we    show    in    Fig.~\ref{fig:model2}    and
Fig.~\ref{fig:model3} a poor and a good  fit to the observed data.  It
is also  worth mentioning  that the  minimum value  of $0.08$  for the
minimal  mass   ratio  parameter  $q_0$  generally   performs  better.
Finally,  among all  IMRDs that  satisfy  that the  angle between  the
theoretical distribution and the observed data is smaller than 13.5 we
find both our reference model ($n(q)=1$) and the most favored model of
\cite{Ducati2011}.

We also  investigated the possible  correlations between the  IMRD and
the       $\alpha_{\mathrm{CE}}$      parameters       by      varying
$\alpha_{\mathrm{CE}}$ from  0.0 to 0.5 for  three representative IMRD
models:  model 1  (flat), model  2 (linearly  increasing) and  model 3
(inversely decreasing).   For each of  these three IMRDs we  find that
the white dwarf effective  temperature distributions and the secondary
spectral type distributions are largely  unaffected by the increase in
$\alpha_{\mathrm{CE}}$.  Moreover, the $\log g$ distribution shows the
same  behaviour   described  in  Section~\ref{sec:res-CE}   (see  also
Fig.\,\ref{fig:logg})     relative     to      the     increase     of
$\alpha_{\mathrm{CE}}$, irrespective of the considered IMRD.

\section{Conclusions}
\label{sec:conc}

We have presented the most detailed population synthesis study to date
of the WD+MS  binary population in the Galactic  disk.  In particular,
our  work  aimed  at  reproducing   the  ensemble  properties  of  the
population of such binaries in the SDSS  data release 12.  To do so we
have    independently   simulated    the   different    WD+MS   binary
sub-populations  observed  by  the  Legacy, BOSS,  SEGUE  and  SEGUE-2
surveys  of  the  SDSS,  taking   special  care  in  implementing  all
observational biases known for each one of them.

We have found that a value of the common envelope efficiency parameter
between 0.2 and 0.3 is compatible with the observational data. This is
true  not only  in  terms of  the percentage  of  WD+MS binaries  that
experienced a common envelope episode,  but also regarding the overall
properties of the distributions  of white dwarf effective temperatures
and  surface  gravities.   This  result is  consistent  with  previous
observational    \citep{Zorotovic2010}   and    population   synthesis
\citep{Toonen2013,Camacho2014,   Zorotovic2014}    analyses   of   the
population of close SDSS WD+MS binaries.

Moreover, we have  adopted 12 different prescriptions  for the initial
mass ratio  distribution of  main sequence  binaries, which  include a
wide variety of shapes.  We have used three distance metric indicators
to compare  the resulting  synthetic M-dwarf  spectral type  and white
dwarf  effective temperature  and surface  gravity distributions  with
their  observed counterparts.  Our results  indicate that  all initial
mass ratio distributions of ascending shape can be excluded.

Finally, it is worth mentioning that the outcome of the simulations is
nearly independent on  the age of the  thin disk (up to a  20 per cent
variation), on the contribution of  WD+MS binaries from the thick disk
and on the fraction  of the internal energy that is  used to eject the
envelope during  the common envelope  episode. All these  findings are
important  by  themselves  and  open  the  possibility  of  using  the
population  of binary  systems  composed  of a  white  dwarf and  main
sequence star to probe the structure  and evolution of our Galaxy, and
also to  elucidate between different  models aimed at  reproducing the
evolution of binary stars.

\section*{Acknowledgements}

This research was  supported by MINECO grant  AYA\-2014-59084-P and by
the  AGAUR.   R.C.   would  like  to  thank  E.   Zamfir  for  helpful
discussions and also acknowledges financial support from the FPI grant
BES-2012-053448.


 \newcommand{\noop}[1]{}

\bsp
\label{lastpage}
\end{document}